\documentclass[aps,pra,pifont,citesort,epsf,twocolumn]{revtex4}

\usepackage{graphicx,amsmath,amssymb}
\usepackage{dcolumn}
\usepackage{bm}

\newcommand{\bk}[1]{\left ( #1\right )}
\newcommand{\eq}[1]{\begin{equation} \newline #1 \end{equation}}
\newcommand{\eqn}[1]{\begin{eqnarray} \newline #1 \end{eqnarray}}
\newcommand{\ee}{&=&}

\newcommand{\hs}{\hspace{0.2cm}}

\newcommand{\ket}[1]{\left |#1\right \rangle}
\newcommand{\braket}[2]{\langle#1|#2\rangle}
\newcommand{\EV}[1]{\left < #1 \right >}

\newcommand{\nn}{\nonumber}

\newcommand{\adag}{\hat{a}^{\dag}}
\newcommand{\anih}{\hat{a}}
\newcommand{\mat}[4]{
\left(\begin{array}{cc}
#1 & #2 \\
#3 & #4
\end{array}\right)}

\newcommand{\matfour}[4]{
\left(\begin{array}{cccc}
#1  \\
#2\\
#3 \\
#4 
\end{array}\right)}
\newcommand{\A}{g^{\frac{\hat{n}}{2}}}
\newcommand{\matfive}[5]{
\left(\begin{array}{ccccc}
#1  \\
#2\\
#3 \\
#4 \\
#5
\end{array}\right)}

\begin{document}

\title{Gaussian Post-selection for Continuous Variable Quantum Cryptography}

\author{Nathan Walk}
\email{walk@physics.uq.edu.au}
\author{Timothy C. Ralph}
\affiliation{
Centre for Quantum Computation and Communication Technology\\
 School of Mathematics and Physics, University of Queensland, St Lucia, Queensland 4072, Australia}
\author{Thomas Symul}
\author{Ping Koy Lam}
\affiliation{Centre for Quantum Computation and Communication Technology\\
Department of Physics, Faculty of Science, Australian National University, ACT 0200, Australia}

\date{\today}

\begin{abstract}
We extend the security proof for continuous variable quantum key distribution protocols using post selection to account for arbitrary eavesdropping attacks by employing the concept of an equivalent protocol where the post-selection is implemented as a series of quantum operations including a virtual distillation.  We introduce a particular `Gaussian' post selection and demonstrate that the security can be calculated using only experimentally accessible quantities.  Finally we explicitly evaluate the performance for the case of a noisy Gaussian channel in the limit of unbounded key length and find improvements over all pre-existing continuous variable protocols in realistic regimes.
\end{abstract}

\maketitle

Quantum key distribution (QKD) is the process of generating a common random key between two parties using a quantum communications protocol.  The power of this method is that the security of the key distribution, and the subsequent communication via a one time pad, is established while making no assumptions about the technological capabilities of a eavesdropper.  This procedure also has the distinction of being the most developed quantum information technology \cite{Scarani:2009p378}.

There are two main flavours of QKD, discrete variable (DV) and continuous variable (CV) which are realised by encoding and then detecting: single photons \cite{Zhao:2008p527} or the quadrature variables of the optical field \cite{Cerf:2007p524} respectively.  The latter kind, which we shall consider here, has the advantage of higher raw bit rates due to the high efficiency and high bandwidth of homodyne detection and ease of integration with existing communications infrastructure.  CV protocols that employ post-selection \cite{Silberhorn:2002p149} - a classical filtering of the measurement results -  enjoy additional advantages in terms of versatility and reconciliation efficiency.

Asymptotic (in the sense of string length) unconditional security for protocols that do not employ post-selection is achieved by first noting the equivalence of an experimentally implemented prepare and measure (P\&M) scheme to an entanglement based (EB) version \cite{Grosshans:2003p526}, followed by the result that for collective attacks security may be bounded from below by assuming the entangled state at the end of the protocol is Gaussian\cite{Navascues:2006p805,GarciaPatron:2006p381} and finally a proof that collective attacks are asymptotically optimal \cite{Renner:2009p1}.  However for protocols using post-selection (PS) this analysis cannot be straightforwardly applied as an equivalent entanglement based picture has yet to be constructed, with security only shown under the assumption of a Gaussian eavesdropping attack \cite{Heid:2007p375}.  

Here we fill this gap and hence demonstrate unconditional security for post-selected CVQKD following the proof method used in \cite{GarciaPatron:2006p381}.  In particular we construct an EB scheme in which the post-selection is replaced by equivalent heralded state transformations.  We show we are able to straightforwardly construct the necessary parameters of this EB scheme from experimental data providing a realistically obtainable bound for the case of collective attacks and hence asymptotically unconditional security.

{\it Security of CVQKD.---}In general one equates each protocol in which: the sender (Alice) prepares an ensemble of quantum states based upon a classical random probability distribution and sends it through the domain of the eavesdropper (Eve) to the recipient (Bob), to an entanglement based scheme in which: Alice prepares an entangled state one half of which is kept and used for a projective measurement; and the other is transmitted to Bob again through Eves domain. The proper choice of the initial entangled state and the projective measurement by Alice allows us to rigorously express any prepare and measure schemes  \cite{Grosshans:2003p526}.  

Bob makes a quadrature measurement upon his received states and then Alice and Bob engage in a reconciliation procedure to correct the errors in their shared classical string. The secret key rate for the entire protocol is then given by 
\eq{K = \beta I(a:b) - I(E:X), \hs X \in \{a,b\}}
where $I(a:b)$ is the Shannon mutual information between classical strings belonging to Alice and Bob at the end of the protocol, $\beta$ is the efficiency of their reconciliation procedure and $I(E:X)$ is the quantum mutual information between either Eve and Bob if considering reverse reconciliation protocols or Eve and Alice if considering direct reconciliation.  

Eve's mutual information is given by the purification of the entangled state before and after Alice or Bob's measurement.  For example the direct reconciliation expression is \cite{Navascues:2006p805,GarciaPatron:2006p381}
\eq{\label{IE}I(E:A) = S(\rho_E) - S(\rho_{E}|a) = S(\rho_{AB}) - S(\rho_{B}|a)}
with the von Neumann entropy given by $S(\rho) = \mathrm{tr}(\rho \log\rho)$ and for the second equality we have used the fact that the overall tripartite state $\ket{ABE}$ is pure.  This quantity is not easy to calculate in general but it has been shown that we may bound the expression from below by analysing a Gaussian \cite{Navascues:2006p805,GarciaPatron:2006p381}, symmetric \cite{Leverrier:2009p390} state with the same first and second moments.  For Gaussian states, the von Neumann entropy is obtained straightforwardly and thus the security of the entire protocol can be characterised entirely by the covariance matrix of the entangled state shared by Alice and Bob.

{\it Equivalent post-selection scheme.---}While reverse reconciliation can be shown to to be secure for arbitrary losses in the absence of noise, for any non-zero amount of excess noise the secure distance is inevitably finite.  One could attempt to address this by increasing the input signal modulation however any imperfection in the reconciliation process ($\beta <1$) means optimising the modulation can also only lead to a finite improvement.  On the other hand direct reconciliation is significantly more tolerant to excess noise but only successful when the channel loss is below 50$\%$ or 3 dB.  

All of these detrimental effects can be improved using post-selection \cite{Silberhorn:2002p149}, a technique in which values in the space of Bob's possible quadrature measurement results and Alice's quadrature encoding are probabilistically re-weighted and only these new distributions are kept to form the key.  Intuitively one would expect this strategy to yield an advantage as the eavesdropper is effectively shut out of Alice and Bob's post-measurement collaboration. This improved performance comes at the price of not being able to directly apply Eq. (\ref{IE}) as this would not allow for Eve's knowledge of Alice and Bob's post-selection.  This can be accounted for as long as one can keep track of the way post-selection by one party influences the state of the other in the equivalent EB scheme which we shall now demonstrate.  
\begin{figure}[htbp]
\begin{center}
\includegraphics[width = 7.8cm]{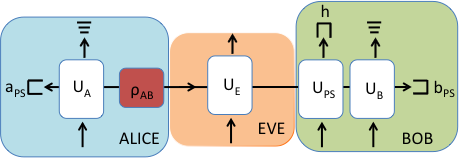}
\caption{Equivalent entanglement based version of a post-selected protocol.  Alice distributes one arm of an entangled state through Eve's domain to Bob and makes a projective measurement $U_A$ (giving classic output a) corresponding to an ensemble of states sent in a prepare and measure scheme. Bob passes his arm first through a device that probabilistically distills entanglement, $U_{PS}$, and then makes a potentially noisy measurement $U_B$ giving classical output $b_{PS}$.  The heralding signal of $U_{PS}$ (h) is given to Eve but the remaining ancillae are kept within the stations of Alice and Bob.} 
\label{ups}
\end{center}
\end{figure}

In general one post selects by applying a weighting function to achieve a new probability distribution in the chosen measurement basis, $p(x)\xrightarrow{\text{PS}} w(x)p(x) = p'(x)$.  The normalisation of $p'(x)$ gives the amount of data retained while transitioning from the initial to the post-selected ensemble.  Alternatively one could apply an appropriate transformation consisting of a unitary acting on the mode in question together with auxiliary mode(s) which are subsequently traced out.  A useful post-selection will correspond in the EB scheme to achieving some amount of distillation of the virtual entanglement, inevitably along with some additional noise.  In this setting, Fig.\ref{ups}, the probabilistic nature of the post selection corresponds to Bob's first operation $U_{PS}$ being a non-deterministic but heralded operation (the distillation) followed by appropriate deterministic unitary interactions corresponding to the noise addition and Bob's final measurement.  

In this picture Eve's additional knowledge of the form of Alice and Bob's post selection is reflected by the state $\rho_{AB}$ in Eq.(\ref{IE}) being identified with the state conditioned on successful heralding of Bob's first operation, i.e. the state after $U_{PS}$ but before $U_B$.  After all ancillae are traced over the outputs of $U_A$ and $U_B$ give classical strings $a$ and $b_{PS}$ which exactly match the experimental results for the post selected ensemble.  Note that we will be considering the secure station scenario where the ancillae remaining within the laboratories of Alice and Bob will not be attributed to the eavesdropper. 

The worst case scenario would correspond to the final state being Gaussian \cite{Navascues:2006p805,GarciaPatron:2006p381} so if one is able to uniquely identify a Gaussian collection of unitaries and ancillae that result in the same measurement statistics then the key rate of that state will be provide a lower bound for the post-selected protocol.  Denoting a successful run of the distillation as the outcome $s$ and the resultant state $\rho_{PS} = \frac{(U_{PS}^s)^{\dag}\rho_{AB}U_{PS}^s}{\mathrm{tr}[(U_{PS}^s)^{\dag}\rho_{AB}U_{PS}^s]}$ one may write,
\eqn{I_{PS}(A:E) = S(\rho_{PS}) - S(\rho_{PS}|a)\nn \\
\leq S(\rho_{PS}^G) - S(\rho_{PS}^G|a),\label{eps}}
where $\rho_{PS}^G$ is a Gaussian state with the same covariance matrix.  This method is applicable to any form of post-selection, however demonstrating that the covariance matrix of the equivalent Gaussian setup can be obtained from measured data is non-trivial and may be more or less experimentally demanding (in terms of the number of measurements) depending upon the particular form of the post-selection.

{\it Gaussian post-selection.---}
\begin{figure}[htbp]
\begin{center}
\includegraphics[width = 7.8cm]{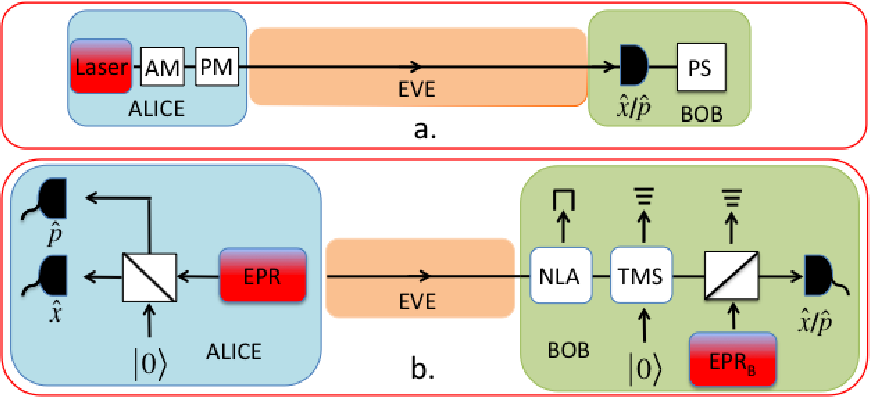}
\caption{Prepare and measure and entanglement based versions of a protocol using Gaussian post-selection. a) P\&M scheme: Alice uses two classical Gaussian strings ($x_A,p_A$) to prepare and transmit an ensemble of coherent states to Bob who homodyne detects and then applies a Gaussian weighting function. b) Equivalent EB scheme:  Alice distributes one arm of an EPR pair and makes a heterodyne measurement obtaining measurement results directly proportional to ($x_A,p_A$). Bob first passes his arm through an NLA, classically amplifies via a vacuum seeded two-mode squeezer (TMS) then mixes his mode with one arm of another entangled pair ($\mathrm{EPR}_B$) on a beamsplitter. He finally homodyne detects and obtains exactly the measurement results from the P\&M scheme.  The heralding signal of the NLA is given to Eve but the unmeasured ancillae are kept within Bob's station.} 
\label{schematic}
\end{center}
\end{figure}
We shall consider a particular P\&M scheme, Fig. \ref{schematic} panel a), in which Alice draws values ($x_A,p_A$) from a bivariate Gaussian of 0 mean and variance $V_A$ and uses these numbers to modulate the vacuum to create an ensemble of coherent states of the form $\ket{x_A + ip_A}$ which she sends to Bob through a quantum channel. Bob uses homodyne detection on his received states, randomly switching between quadratures given by $\hat{x} = \anih + \adag$ and $\hat{p} = i(\adag - \anih)$ where we have normalised the vacuum noise to unity. Bob then filters his results with the goal of selecting an ensemble which is a Gaussian distribution with a certain target variance $V_{PS}$.  For the most common case of a Gaussian channel Bob's input distribution is another Gaussian of variance $V_B$ and the appropriate weighting function would look like $w(x) = \sqrt{V_B/V_{PS}} \exp(-x^2(1/V_{PS} - 1/V_B))$.  In the relevant case $V_{PS}>V_B$ this function is convex so in order to arrive at a proper probability distribution we will choose some endpoints $\pm \Delta$, renormalise the function to the value at this point and set all values outside this range to unity.  For a Gaussian input state the exact filter function is
\eq{W(x) = N\bk{1 + \bk{\frac{w(x)}{w(\Delta)} - 1}\bk{\Theta(x+\Delta) - \Theta(x - \Delta)}}\label{W2}}
where $\Theta(x)$ is the Heaviside step function and the fraction of data kept is the re-normalisation $N$.
When $\Delta$ is 0 this operation is the identity. As $\Delta\rightarrow\infty$ it results in a Gaussian distribution of variance $V_{PS}$ and in between give a slightly non-Gaussian state with variance $V_B<V<V_{PS}$.  
Finally Alice and Bob publicly announce a subset of their data to characterise the covariance matrix on both the initial and post-selected ensemble and, if secure, engage in reconciliation and privacy amplification to distill a completely secure key.  Notice that although the weight function is smooth instead of hard edged it is determined entirely by Bob and the only information he sends to Alice is a `keep or reject' signal.

The equivalent entanglement based scheme, Fig. \ref{schematic} panel b), involves Alice preparing a two-mode squeezed vacuum or EPR state, one mode of which she keeps and measures, the other being transmitted to Bob.  Alice's makes a heterodyne detection whereas Bob's measurement, depending upon the target variance, decomposes into a combination of a noiseless amplification/distillation followed by classical amplification and finally some additional noise.  The necessary Gaussian entanglement distillation is achieved via the noiseless linear amplifier (NLA) \cite{proc-disc-2009,Xiang:2010p1449} with the classical amplifier and additional noise corresponding to a two-mode squeezer with vacuum ancilla and a beamsplitter with an EPR pair ancilla respectively. If we can uniquely characterise Gaussian operations that perform the necessary transformations from the transmitted to the post-selected ensemble at the level of the covariance matrix then we can apply the above proof and determine the security.  

To illustrate this method we evaluate performance for the most common case, the noisy Gaussian channel. Such a channel is completely parameterised by it's transmission $T$ and excess thermal noise $\xi$ \cite{fnote}. One can calculate the action of an NLA on an EPR state sent through a general Gaussian channel \cite{supp,Blandino:2012p4960} with the result being an effective protocol where stronger entanglement was distributed through a channel with less loss but greater excess noise, leading to an overall advantage.  Inverting the relationship for the effective entanglement generated gives a relationship,
\eq{g= 1+\frac{2 (V_A^{PS}-V_A)}{T (V_A (2+V_A^{PS}-\xi )+V_A^{PS}\xi )}}
 that uniquely identifies the gain of the NLA based only upon the measured channel parameters and Alice's modulation variance before and after after post-selection.  
Bob and Alice's other operations are just beamsplitters and two-mode squeezing, their effect on the covariance matrix being given by the appropriate symplectic transformations \cite{Weedbrook:2012p5160}.   Given Alice and Bob's measurement of the covariance matrix before and after the post-selection straightforward algebra allows us to characterise all parameters in Fig.\ref{schematic} and thus unconditionally bound Eve's information via Eq.(\ref{eps}).  For this form of post-selection, the resultant distributions turn out to be extremely close to Gaussian, so we have the option of making use of recent improvements in the reconciliation efficiency Gaussian variables \cite{Jouguet:2011p5057} as well as the sign encoding of \cite{Silberhorn:2002p149} and we will choose the former case here. See  \cite{supp} for a detailed calculation of these quantities and the secret key rate.  The crucial tradeoff in this scheme is between a large post-selection to improve the effective channel and the proportion of measurement results that are discarded. 

\begin{figure}[htbp]
\begin{center}
\includegraphics[width = 8cm]{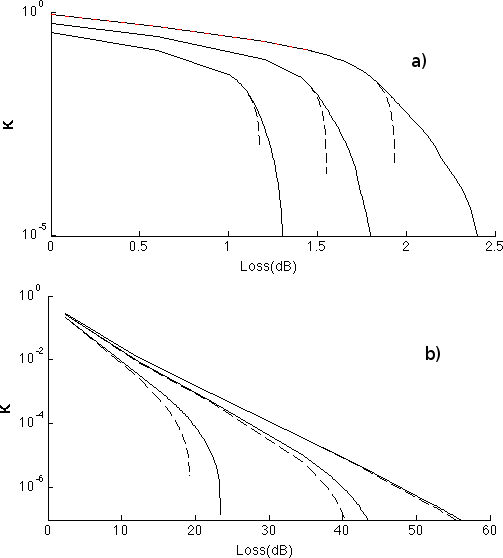}
\caption{\label{K} Improvement in secret key rates due to Gaussian post-selection.  a) Direct reconciliation with post-selection (solid lines) and without (dashed lines) as a function of loss for $\xi = \{0.1, 0.2, 0.3\}$ with decreasing key rate. b) Reverse reconciliation with post-selection (solid lines) and without (dashed lines) as a function of loss for $\xi = \{0.02, 0.03, 0.05\}$ with decreasing key rate.  For all plots $\beta = 0.9$.}
\end{center}
\end{figure}

We plot the key rate as a function of distance of a coherent state homodyne protocol, Fig.\ref{K}, for both direct and reverse reconciliation along with the case without post-selection for comparison.  In all plots the reconciliation efficiency is taken to be a constant value of $\beta = 0.9$ and for each point Alice's modulation variance is independently optimised for each protocol along with the parameters ($\Delta$,$V_{PS}$) for the post-selected scheme. Furthermore for the case without post-selection we included the potential to optimise over a deterministic addition of noise as outlined in \cite{GarciaPatron:2009p3479}. For these realistic experimental parameters the post-selection protocol allows for secure key generation over long distances (in combination with RR) and for greater excess noise (in combination with DR) than any previous coherent state protocol.  Finally the presence of an NLA in the effective Gaussian circuit leads one to compare these results with those of \cite{Blandino:2012p4960}. We find that almost all of the improvements shown there are recovered by our classical post-processing scheme.  This leads one to conclude that an NLA placed just before Bob's detectors will not lead to a benefit for QKD over and above that of a post-selection scheme.

{\it Conclusions.---}

In conclusion we have shown that post-selection based CVQKD is secure for arbitrary collective attacks, and thus asymptotically secure for all attacks.  This was achieved by identifying an entanglement based scheme that correctly reflects the post-selected ensemble that is used in the final key generation.  Furthermore  results for a particular Gaussian form of post-selection show improvements in performance over all previous coherent state protocols for certain relevant combinations of loss and noise.  Avenues for further work, include the investigation of other post-selection filters with a view to proving which is optimal, the incorporation of finite-size effects and the combination of post-selection with other protocols to identify the optimal technique for a given scenario.

The authors would like to thank Norbert L\"utkenhaus, Christian Weedbrook, Anthony Leverrier, Remi Blandino and Andrew Lance for helpful discussions.  This research was conducted by the Australian Research Council Centre of Excellence for Quantum Computation and Communication Technology (Project number CE11000102).

We have recently become aware of independent work on virtual noiseless amplification in CVQKD \cite{Fiurasek:2012p5315}.

\bibliographystyle{apsrev}
\bibliography{Gaussian_post_selection_for_CVQKD_arxiv2}

\begin{thebibliography}{21}
\expandafter\ifx\csname natexlab\endcsname\relax\def\natexlab#1{#1}\fi
\expandafter\ifx\csname bibnamefont\endcsname\relax
  \def\bibnamefont#1{#1}\fi
\expandafter\ifx\csname bibfnamefont\endcsname\relax
  \def\bibfnamefont#1{#1}\fi
\expandafter\ifx\csname citenamefont\endcsname\relax
  \def\citenamefont#1{#1}\fi
\expandafter\ifx\csname url\endcsname\relax
  \def\url#1{\texttt{#1}}\fi
\expandafter\ifx\csname urlprefix\endcsname\relax\def\urlprefix{URL }\fi
\providecommand{\bibinfo}[2]{#2}
\providecommand{\eprint}[2][]{\url{#2}}

\bibitem[{\citenamefont{Scarani et~al.}(2009)\citenamefont{Scarani,
  Bechmann-Pasquinucci, Cerf, Du{\v s}ek, L{\"u}tkenhaus, and
  Peev}}]{Scarani:2009p378}
\bibinfo{author}{\bibfnamefont{V.}~\bibnamefont{Scarani}},
  \bibinfo{author}{\bibfnamefont{H.}~\bibnamefont{Bechmann-Pasquinucci}},
  \bibinfo{author}{\bibfnamefont{N.}~\bibnamefont{Cerf}},
  \bibinfo{author}{\bibfnamefont{M.}~\bibnamefont{Du{\v s}ek}},
  \bibinfo{author}{\bibfnamefont{N.}~\bibnamefont{L{\"u}tkenhaus}},
  \bibnamefont{and} \bibinfo{author}{\bibfnamefont{M.}~\bibnamefont{Peev}},
  \bibinfo{journal}{Rev. Mod. Phys.} \textbf{\bibinfo{volume}{81}},
  \bibinfo{pages}{1301} (\bibinfo{year}{2009}).

\bibitem[{\citenamefont{Zhao and Lo}(2008)}]{Zhao:2008p527}
\bibinfo{author}{\bibfnamefont{Y.}~\bibnamefont{Zhao}} \bibnamefont{and}
  \bibinfo{author}{\bibfnamefont{H.-K.} \bibnamefont{Lo}},
  \bibinfo{journal}{arXiv} \textbf{\bibinfo{volume}{quant-ph}}
  (\bibinfo{year}{2008}), \eprint{0803.2507v4}.

\bibitem[{\citenamefont{Cerf and Grangier}(2007)}]{Cerf:2007p524}
\bibinfo{author}{\bibfnamefont{N.}~\bibnamefont{Cerf}} \bibnamefont{and}
  \bibinfo{author}{\bibfnamefont{P.}~\bibnamefont{Grangier}},
  \bibinfo{journal}{JOSA B} \textbf{\bibinfo{volume}{24}}, \bibinfo{pages}{324}
  (\bibinfo{year}{2007}).

\bibitem[{\citenamefont{Silberhorn et~al.}(2002)\citenamefont{Silberhorn,
  Ralph, L{\"u}tkenhaus, and Leuchs}}]{Silberhorn:2002p149}
\bibinfo{author}{\bibfnamefont{C.}~\bibnamefont{Silberhorn}},
  \bibinfo{author}{\bibfnamefont{T.}~\bibnamefont{Ralph}},
  \bibinfo{author}{\bibfnamefont{N.}~\bibnamefont{L{\"u}tkenhaus}},
  \bibnamefont{and} \bibinfo{author}{\bibfnamefont{G.}~\bibnamefont{Leuchs}},
  \bibinfo{journal}{Phys. Rev. Lett.} \textbf{\bibinfo{volume}{89}},
  \bibinfo{pages}{167901} (\bibinfo{year}{2002}).

\bibitem[{\citenamefont{Grosshans et~al.}(2003)\citenamefont{Grosshans, Cerf,
  Grangier, Wenger, and Tualle-Brouri}}]{Grosshans:2003p526}
\bibinfo{author}{\bibfnamefont{F.}~\bibnamefont{Grosshans}},
  \bibinfo{author}{\bibfnamefont{N.}~\bibnamefont{Cerf}},
  \bibinfo{author}{\bibfnamefont{P.}~\bibnamefont{Grangier}},
  \bibinfo{author}{\bibfnamefont{J.}~\bibnamefont{Wenger}}, \bibnamefont{and}
  \bibinfo{author}{\bibfnamefont{R.}~\bibnamefont{Tualle-Brouri}},
  \bibinfo{journal}{Quantum Inf. Comput.} \textbf{\bibinfo{volume}{3}},
  \bibinfo{pages}{535} (\bibinfo{year}{2003}).

\bibitem[{\citenamefont{Navascu{\'e}s et~al.}(2006)\citenamefont{Navascu{\'e}s,
  Grosshans, and Ac{\'\i}n}}]{Navascues:2006p805}
\bibinfo{author}{\bibfnamefont{M.}~\bibnamefont{Navascu{\'e}s}},
  \bibinfo{author}{\bibfnamefont{F.}~\bibnamefont{Grosshans}},
  \bibnamefont{and}
  \bibinfo{author}{\bibfnamefont{A.}~\bibnamefont{Ac{\'\i}n}},
  \bibinfo{journal}{Phys. Rev. Lett.} \textbf{\bibinfo{volume}{97}},
  \bibinfo{pages}{190502} (\bibinfo{year}{2006}).

\bibitem[{\citenamefont{Garc{\'\i}a-Patr{\'o}n and
  Cerf}(2006)}]{GarciaPatron:2006p381}
\bibinfo{author}{\bibfnamefont{R.}~\bibnamefont{Garc{\'\i}a-Patr{\'o}n}}
  \bibnamefont{and} \bibinfo{author}{\bibfnamefont{N.}~\bibnamefont{Cerf}},
  \bibinfo{journal}{Phys. Rev. Lett.} \textbf{\bibinfo{volume}{97}},
  \bibinfo{pages}{190503} (\bibinfo{year}{2006}).

\bibitem[{\citenamefont{Renner and Cirac}(2009)}]{Renner:2009p1}
\bibinfo{author}{\bibfnamefont{R.}~\bibnamefont{Renner}} \bibnamefont{and}
  \bibinfo{author}{\bibfnamefont{J.}~\bibnamefont{Cirac}},
  \bibinfo{journal}{Phys. Rev. Lett.} \textbf{\bibinfo{volume}{102}},
  \bibinfo{pages}{110504} (\bibinfo{year}{2009}).

\bibitem[{\citenamefont{Heid and L{\"u}tkenhaus}(2007)}]{Heid:2007p375}
\bibinfo{author}{\bibfnamefont{M.}~\bibnamefont{Heid}} \bibnamefont{and}
  \bibinfo{author}{\bibfnamefont{N.}~\bibnamefont{L{\"u}tkenhaus}},
  \bibinfo{journal}{Phys. Rev. A} \textbf{\bibinfo{volume}{76}},
  \bibinfo{pages}{022313} (\bibinfo{year}{2007}).

\bibitem[{\citenamefont{Leverrier et~al.}(2009)\citenamefont{Leverrier, Karpov,
  Grangier, and Cerf}}]{Leverrier:2009p390}
\bibinfo{author}{\bibfnamefont{A.}~\bibnamefont{Leverrier}},
  \bibinfo{author}{\bibfnamefont{E.}~\bibnamefont{Karpov}},
  \bibinfo{author}{\bibfnamefont{P.}~\bibnamefont{Grangier}}, \bibnamefont{and}
  \bibinfo{author}{\bibfnamefont{N.}~\bibnamefont{Cerf}}, \bibinfo{journal}{New
  Journal of Physics} \textbf{\bibinfo{volume}{11}}, \bibinfo{pages}{115009}
  (\bibinfo{year}{2009}).

\bibitem[{\citenamefont{Ralph and Lund}(2009)}]{proc-disc-2009}
\bibinfo{author}{\bibfnamefont{T.}~\bibnamefont{Ralph}} \bibnamefont{and}
  \bibinfo{author}{\bibfnamefont{A.}~\bibnamefont{Lund}},
  \bibinfo{journal}{{\it Quantum Communication Measurement and Computing
  Proceedings of 9th International Conference}} p. \bibinfo{pages}{155}
  (\bibinfo{year}{2009}).

\bibitem[{\citenamefont{Xiang et~al.}(2010)\citenamefont{Xiang, Ralph, Lund,
  Walk, and Pryde}}]{Xiang:2010p1449}
\bibinfo{author}{\bibfnamefont{G.}~\bibnamefont{Xiang}},
  \bibinfo{author}{\bibfnamefont{T.}~\bibnamefont{Ralph}},
  \bibinfo{author}{\bibfnamefont{A.}~\bibnamefont{Lund}},
  \bibinfo{author}{\bibfnamefont{N.}~\bibnamefont{Walk}}, \bibnamefont{and}
  \bibinfo{author}{\bibfnamefont{G.}~\bibnamefont{Pryde}},
  \bibinfo{journal}{Nature Photonics} \textbf{\bibinfo{volume}{4}},
  \bibinfo{pages}{316} (\bibinfo{year}{2010}).

\bibitem[{fno()}]{fnote}
\bibinfo{note}{Note that this is the excess noise referred to the input as
  opposed to the quantity $T\xi$ that would be directly measured by Bob.}

\bibitem[{sup()}]{supp}
\bibinfo{note}{Supplementary Material}.

\bibitem[{\citenamefont{Blandino et~al.}(2012)\citenamefont{Blandino,
  Leverrier, Barbieri, Etesse, Grangier, and
  Tualle-Brouri}}]{Blandino:2012p4960}
\bibinfo{author}{\bibfnamefont{R.}~\bibnamefont{Blandino}},
  \bibinfo{author}{\bibfnamefont{A.}~\bibnamefont{Leverrier}},
  \bibinfo{author}{\bibfnamefont{M.}~\bibnamefont{Barbieri}},
  \bibinfo{author}{\bibfnamefont{J.}~\bibnamefont{Etesse}},
  \bibinfo{author}{\bibfnamefont{P.}~\bibnamefont{Grangier}}, \bibnamefont{and}
  \bibinfo{author}{\bibfnamefont{R.}~\bibnamefont{Tualle-Brouri}},
  \bibinfo{journal}{arXiv} \textbf{\bibinfo{volume}{quant-ph}}
  (\bibinfo{year}{2012}), \eprint{1205.0959v1}.

\bibitem[{\citenamefont{Weedbrook et~al.}(2012)\citenamefont{Weedbrook,
  Pirandola, Garc{\'\i}a-Patr{\'o}n, Cerf, Ralph, Shapiro, and
  Lloyd}}]{Weedbrook:2012p5160}
\bibinfo{author}{\bibfnamefont{C.}~\bibnamefont{Weedbrook}},
  \bibinfo{author}{\bibfnamefont{S.}~\bibnamefont{Pirandola}},
  \bibinfo{author}{\bibfnamefont{R.}~\bibnamefont{Garc{\'\i}a-Patr{\'o}n}},
  \bibinfo{author}{\bibfnamefont{N.}~\bibnamefont{Cerf}},
  \bibinfo{author}{\bibfnamefont{T.}~\bibnamefont{Ralph}},
  \bibinfo{author}{\bibfnamefont{J.}~\bibnamefont{Shapiro}}, \bibnamefont{and}
  \bibinfo{author}{\bibfnamefont{S.}~\bibnamefont{Lloyd}},
  \bibinfo{journal}{Rev. Mod. Phys.} \textbf{\bibinfo{volume}{84}},
  \bibinfo{pages}{621} (\bibinfo{year}{2012}).

\bibitem[{\citenamefont{Jouguet et~al.}(2011)\citenamefont{Jouguet,
  Kunz-Jacques, and Leverrier}}]{Jouguet:2011p5057}
\bibinfo{author}{\bibfnamefont{P.}~\bibnamefont{Jouguet}},
  \bibinfo{author}{\bibfnamefont{S.}~\bibnamefont{Kunz-Jacques}},
  \bibnamefont{and}
  \bibinfo{author}{\bibfnamefont{A.}~\bibnamefont{Leverrier}},
  \bibinfo{journal}{Physical Review A} \textbf{\bibinfo{volume}{84}},
  \bibinfo{pages}{062317} (\bibinfo{year}{2011}).

\bibitem[{\citenamefont{Garc{\'\i}a-Patr{\'o}n and
  Cerf}(2009)}]{GarciaPatron:2009p3479}
\bibinfo{author}{\bibfnamefont{R.}~\bibnamefont{Garc{\'\i}a-Patr{\'o}n}}
  \bibnamefont{and} \bibinfo{author}{\bibfnamefont{N.}~\bibnamefont{Cerf}},
  \bibinfo{journal}{Physical Review Letters} \textbf{\bibinfo{volume}{102}},
  \bibinfo{pages}{130501} (\bibinfo{year}{2009}).

\bibitem[{\citenamefont{Fiurasek and Cerf}(2012)}]{Fiurasek:2012p5315}
\bibinfo{author}{\bibfnamefont{J.}~\bibnamefont{Fiurasek}} \bibnamefont{and}
  \bibinfo{author}{\bibfnamefont{N.~J.} \bibnamefont{Cerf}},
  \bibinfo{journal}{arXiv} \textbf{\bibinfo{volume}{quant-ph}}
  (\bibinfo{year}{2012}), \eprint{1205.6933v1}.

\bibitem[{\citenamefont{Leverrier and Grangier}(2011)}]{Leverrier:2011p1357}
\bibinfo{author}{\bibfnamefont{A.}~\bibnamefont{Leverrier}} \bibnamefont{and}
  \bibinfo{author}{\bibfnamefont{P.}~\bibnamefont{Grangier}},
  \bibinfo{journal}{Phy. Rev. A} \textbf{\bibinfo{volume}{83}},
  \bibinfo{pages}{042312} (\bibinfo{year}{2011}).

\bibitem[{\citenamefont{Serafini}(2006)}]{Serafini:2006p5052}
\bibinfo{author}{\bibfnamefont{A.}~\bibnamefont{Serafini}},
  \bibinfo{journal}{Physical Review Letters} \textbf{\bibinfo{volume}{96}},
  \bibinfo{pages}{110402} (\bibinfo{year}{2006}).

\end{thebibliography}
\appendix
\section{Parameter Estimation}
A crucial step in all cryptographic schemes is the estimation of parameters which form the basis for quantitative analysis of security. A further complication is that for many QKD protocols, including the scheme investigated in this Letter, security is actually calculated on a related entanglement based version. Thus it is important to establish that the parameters measured in the experimental implementation are sufficient to characterise the appropriate entangled state.

From a purely theoretical point of view, an appropriate entangled state would be any two-mode state such that Alice's projective measurement results in exactly the same key-generating states as seen by Bob and Eve. This means that for any given experiment there exist multiple options for an equivalent entangled state. However some are much more attractive than others for the purposes of parameter estimation. Here we shall first go through the procedure for estimating the necessary quantities using only experimentally accessible data for the canonical Gaussian protocols and then show that a similar procedure exists for the scheme outlined in this letter.
In the all Gaussian CVQKD protocols the procedure is relatively straightforward. One is interested in estimating a covariance matrix (CM) of the form
\eqn{\gamma_{AB}\label{sym} = \mat{a\hs\mathbb{I}_2}{c \hs \sigma_z}{c\hs\sigma_z}{b\hs\mathbb{I}_2}}
where $\mathbb{I}_2 = [1,0;0,1]$, $\sigma_z = [1,0;0,-1]$.  
The block form is justified with the use of symmetry arguments as outlined in \cite{Leverrier:2009p390,Leverrier:2011p1357} and thus one must only know 3 parameters. The experimentally available information consists of 2 classical strings: the quadrature displacements of the coherent states created by Alice and the quadrature measurements made by Bob. From these one can compute 2 variances and a covariance which will give us the 3 required parameters as shown below. In all following discussion the vacuum noise is normalised to unity.  The canonical choice of entangled purification for the Gaussian ensemble of coherent states sent by Alice is an EPR pair  \cite{Grosshans:2003p526} which has a covariance matrix of the form \ref{sym} with entries $a = b = V_A+1$ and $c = \sqrt{V_A^2 + 2V_A}$ where $V_A$ is Alice's modulation variance in the P\&M scheme.  

After transmission Alice and Bob will measure the CM of the de-cohered EPR and interpreting these as corresponding to an effective Gaussian channel paramaterised by $T$ and $\xi$ can express the entries as $a = V_A+1$, $b = TV_A + T\xi +1$ and $c = \sqrt{T(V_A^2+2V_A)}$.  

Up to this point Alice and Bob follow exactly the same procedure for our new protocol.  Then, after they apply the post-selection filter they repeat this process to measure another CM, $\gamma_{PS}$, with entries given by  $a = V_A^{PS}+1$, $b = T^{PS}V_A^{PS} + T^{PS}\xi^{PS} +1$ and $c = \sqrt{T^{PS}((V_A^{PS})^2+2V_A^{PS})}$.  Note that although we write down effective channel parameters the elements of the CM are directly measured and not assumed to take some form based upon the channel parameters.

We now wish to determine the Gaussian operations within Bob's station that will result in the same statistics.  These are an NLA gain $g$, a parametric amplification parameter $\eta$ and an EPR of variance $N_B$ injected through a beamsplitter of transmission $T_B$. Inverting the expressions for the NLA derived below we can solve for the gain to find,
\eq{g= 1+\frac{2 (V_A^{PS}-V_A)}{T (V_A (2+V_A^{PS}-\xi )+V_A^{PS}\xi )}.}
Substituting this gain into the transformations above we deduce a third covariance matrix of the state after the NLA, $\gamma_{NLA}$, with entries $a_{NLA},b_{NLA},c_{NLA}$ of the same form as above . In general the Gaussian post-selection does not only simulate an NLA but acts to further amplify the state and then add noise.  Both these operations are equivalent to unitary interactions with ancillae which are subsequently traced out.  The amplification corresponds to interacting the incoming mode with the vacuum in a two-mode squeezer and the thermal noise is equivalent to mixing the incoming mode with one arm of an EPR pair on a beamsplitter.  These operations between these last two covariance matrices can thus be compactly expressed via symplectic transformations,
\eq{\label{gtrans}\gamma_{PS} = {\bf BS}^T\bk{T_B}{\bf S}^T(\eta)\gamma_{NLA}{\bf S}(\eta){\bf BS}\bk{T_B}}
where 
\eqn{\nn {\bf S}(\eta) \ee  \mat{\sqrt{\eta} \hs\mathbb{I}_2}{\sqrt{\eta-1}\hs\sigma_z}{\sqrt{\eta-1}\hs\sigma_z}{\sqrt{\eta}\hs\mathbb{I}_2}, \\
{\bf BS}(T) \ee \mat{\sqrt{T}\hs\mathbb{I}_2}{\sqrt{1-T}\hs\mathbb{I}_2}{-\sqrt{1-T}\hs\mathbb{I}_2}{\sqrt{T}\hs\mathbb{I}_2}\label{S}}
are the symplectic transformations for a two-mode squeezer and a beamsplitter respectively.  
Substituting $\gamma_{NLA}$ and $\gamma_{PS}$ leaves us with a matrix equation that can be solved for $T_B$, $\eta$ and $N_B$.  The necessary relationships are,
\eqn{\nn T_B(N_B+1)-N_B \ee b_{PS} - \frac{c_{PS}^2}{c_{NLA}^2}(1+b_{NLA})\\
\eta \ee \frac{c_{PS}^2}{c_{NLA}^2 T_B}}
Strictly speaking there is a degeneracy between these parameters in that there are combinations of values the three, however any combination consistent with the output statistics result in the same secret key rate.  Thus we can construct all the necessary parameters to calculate the secret key rate based only upon the directly measured elements of the covariance matrix before and after post-selection.
\section{NLA: Application to CVQKD}
As described above the ensemble that Alice sends to Bob is equivalent to the preparation of an EPR pair of variance $V = V_A+1$ which in turn can be created by mixing appropriately squeezed states on a 50:50 beamsplitter. Furthermore transmission through a Gaussian channel of parameters $T$ and $\xi$ is also equivalent to interaction on a beamsplitter of transmission $T$ with a second EPR created by Eve with variance given by $N_E = \frac{1-T+T\xi}{1-T}$.  Thus we can calculate the effect of on NLA on a CVQKD system through a Gaussian channel by mixing two appropriate EPR pairs and then applying the NLA to Bob's mode.  

The initial EPR states of Alice and Eve can be written in the $\hat{x}$ quadrature basis,
\eqn{\nn\ket{AB} \ee \frac{1}{\sqrt{2\pi}} \int d^2x_i\hs e^{-\frac{1}{8}\frac{(x_A - x_B)^2}{V_S}} e^{-\frac{1}{8} (x_A + x_B)^2V_S} \ket{x_A}\ket{x_B}\\
\ket{E_1E_2} \ee\frac{1}{\sqrt{2\pi}} \int d^2x_i \hs e^{-\frac{1}{8}\frac{(x_{E_1} - x_{E_2})^2}{V_\xi}} e^{-\frac{1}{8} (x_{E_1} + x_{E_2})^2V_\xi} \ket{x_{E_2}}\ket{x_{E_1}}\nn}
where the squeezing variances are related to the $P\&M$ parameters by $V_S = e^{\mathrm{acosh}(V_A+1)}$ and $V_\xi = e^{\mathrm{acosh}(N_E)}$ and in all integrals $dx_i$ to to be taken to run over all mode variables present.  
These modes $B$ and $E_1$ are then mixed on a beamsplitter of transmission $T$ with one output being sent to Bob and the other retained by Eve.  If we transform to the Hilbert space belonging to Alice Bob and Eve at the end of the protocol the appropriate change of variables allows us to write the final shared state as,
\eqn{\label{ABE} \ket{ABE} \ee \frac{1}{2\pi} \int d^2x_i\hs  \psi(x_i) \ket{x_A}\ket{x_{E_2}}\ket{x_{E_1}}\ket{x_{E_2}}}
where
\begin{widetext}
\eq{\psi(x_i) = \frac{1}{2 \pi }e^{-\frac{\left(x_A+\sqrt{T} x_B+\sqrt{1-T} x_{e_1}\right){}^2}{8 V_S}-\frac{1}{4} V_S \left(-\frac{x_A}{\sqrt{2}}+\frac{\sqrt{T} x_B}{\sqrt{2}}+\frac{\sqrt{1-T} x_{E_1}}{\sqrt{2}}\right){}^2-\frac{1}{16} V_{\xi } \left(-\sqrt{2} \sqrt{1-T} x_B+\sqrt{2} \sqrt{T} x_{E_1}-\sqrt{2} x_{E_2}\right){}^2-\frac{\left(-\sqrt{2} \sqrt{1-T} x_B+\sqrt{2} \sqrt{T} x_{E_1}+\sqrt{2} x_{E_2}\right){}^2}{16 V_{\xi }}}}
\end{widetext}

We consider a non-deterministic amplification operator of a fixed intensity gain $g$ that acts upon Fock states according to $g^{\frac{\hat{n}}{2}}\ket{n} = g^{\frac{n}{2}}\ket{n}$.  For compactness we will suppress all kets except for Bob's mode where the NLA is to be applied.  Applying the NLA to our state we insert a number state resolution of the identity, apply the NLA and then transform back to the quadrature basis via another resolution of the identity,
\eqn{\ket{ABE} \ee\nn \int dx \hs \psi(x)\A\ket{x}\\
\ee \int dx \hs \psi(x) \sum_{n=0}^\infty \A\ket{n}\braket{n}{x}\nn\\
\ee  \int dx \hs \psi(x) \sum_{n=0}^\infty g^{\frac{n}{2}}\ket{n}\braket{n}{x}\nn\\
\ee \int dx dy \hs \psi(x) \sum_{n=0}^\infty g^{\frac{n}{2}} \braket{n}{x} \braket{y}{n} \sqrt{g}^{n} \ket{y} \label{ABE2}}
For the summand we use the definition of a correctly normalised overlap between a number state and a quadrature state,
\eq{\nn \braket{x}{n} = \frac{e^{-\frac{x^2}{4}}H_n\bk{\frac{x}{\sqrt{2}}}}{(2\pi)^{\frac{1}{4}} \sqrt{2^n n!}}}
where $H_n(x) = (-1)^ne^{x^2}\frac{d^n}{dx^n}e^{-x^2}$ are the Hermite polynomials, to arrive at,

\eqn{\sum_{n=0}^\infty \braket{n}{x} \braket{y}{n} \sqrt{g}^{n}\ee\nn \sum_{n=0}^\infty  \frac{e^{-\frac{x^2}{4} - \frac{y^2}{4} }H_n\bk{\frac{x}{\sqrt{2}}}H_n\bk{\frac{y}{\sqrt{2}}} \sqrt{g}^{n}}{\sqrt{2\pi}2^n n!}}
Finally we make use of the following identity,
\eq{\sum_{n=0}^\infty \frac{H_n(x)H_n(y)}{n!}\bk{\frac{u}{2}}^n = \frac{1}{\sqrt{1-u^2}} e^{\bk{\frac{2uxy}{u+1} - \frac{u^2}{u^2-1}(x-y)^2}}\label{id}}
Substituting Eq.(\ref{id}) into Eq.(\ref{ABE2}) and then Eq.(\ref{ABE}) allows us to compute the elements of Alice and Bob's new covariance matrix via lengthy but straightforward integration.  The expressions for the elements themselves are cumbersome but the resultant expressions can be solved to give effective parameters that characterise an effective EPR sent through an effective channel.  To facilitate easy comparison with the previous work \cite{proc-disc-2009,Xiang:2010p1449} we will rewrite the effective modulation variance as an effective EPR entanglement parameter related by $\chi = \sqrt{\frac{V_A}{V_A+2}}$.  These effective parameters are given by,
\eqn{\chi^{NLA} \ee\nn \sqrt{1+\frac{2T(1-g))}{g T \xi -T \xi -2)}}\chi\\
T^{NLA} \ee\nn \frac{4 g T}{(-2 + (-1 + g) T (-2 + \xi)) (-2 + (-1 + g) T \xi)}\\
 \xi^{NLA} \ee \xi  - \xi  T(\xi -2)(g-1)/2 \label{effparam}}
An important sanity check is that in the limit $g\rightarrow 1$ these expressions simply return the input values and for $\xi = 0$ correspond to the results of  \cite{proc-disc-2009,Xiang:2010p1449}.  These results have been independently derived in \cite{Blandino:2012p4960} by a different method, namely considering the action of the NLA on the thermal state that is Bob's output.

\section{Calculation of Key rate}
Now we have all the tools necessary to derive the key rate for the case of a noisy Gaussian channel utilising Gaussian post-selection. This is given by the difference in the mutual information between the two legitimate parties, and the reference party and the eavesdropper.  First we need to calculate the covariance matrix of the post-selected ensemble.  

We recall that in the EB scheme if Alice were to make a heterodyne measurement of $x_A, p_A$ upon her arm of the EPR she would find a Gaussian distribution distribution with variance $V = \frac{V_A+2}{2}$ and would project Bob's arm into a coherent state centered at $\sqrt{\frac{V_A}{2+V_A}}(x_A,p_A)$.  After transmission Bob's receives displaced thermal states of variance $V_B = TV_A + T\xi +1$. Writing $P(x_A,x_B) = P(x_A)P(x_B|x_A)$ we have for the conditional probability distribution,
\eq{P(x_A,x_B) = \frac{e^{-\frac{x_A^2}{2+V_A}-\frac{\left(-\sqrt{\frac{T V_A}{2+V_A}} x_A+x_B\right){}^2}{2 (1+T \xi )}}}{\sqrt{2} \pi  \sqrt{1+T \xi } \sqrt{2+V_A}}}
As explained earlier Bob then applies a weighting function given by
\eq{W(x) = N\bk{1 + \bk{\frac{w(x)}{w(\Delta)} - 1}\bk{\Theta(x+\Delta) - \Theta(x - \Delta)}}\label{W2}}
where  $w(x) = \sqrt{V_B/V_{PS}} \exp(-x^2(1/V_{PS} - 1/V_B))$, $\Theta(x)$ is the Heaviside step function, $\Delta$ is a cutoff point and N is the renormalisation.  Thus the post-selected joint probability distribution is given by,
\eq{P_{PS}(x_A,x_B) = P(x_A,x_B)W(x_B)} 
The fraction of data kept is the re-normalisation $N^{-1} = \int dx_A dx_B P(x_A,x_B)W(x_B)$.  For the post-selected covariance matrix the  necessary moments, $M\bk{x_A,x_B} = x_A^2,x_Ax_B$ etc., are found by evaluating the integral
\eqn{\EV{M(x_A,x_B)} = \int dx_Adx_B \hs P_{PS}(x_A,x_B)M\bk{x_A,x_B} \label{mom}}

\subsection{Alice and Bob Mutual Information}
The maximum information Alice and Bob can extract from two correlated variables with joint probability distribution $P(x_A,x_B)$ is given by Shannon's formula,
\eqn{I(a:b) = P(x_A,x_B)\log\bk{\frac{P(x_A,x_B)}{P(x_A)P(x_B)}}\label{shannon}}
where $P(x_A),P(x_B)$ are the marginal probability distributions and is the same for reverse and direct reconciliation.  In the case of Gaussian distributions this would reduce to the familiar formula $I_G(a:b) = \frac{1}{2}\log(1+SNR)$.  As outlined in the paper for a target variance $V_{PS}>V_B$  there will be some values of $\Delta$ for which our distributions will diverge from Gaussian but this non-Gaussianity turns out to be very small.  
Finding efficient reconciliation codes to extract the maximum information is non-trivial but as the distributions that result from this protocol are extremely close to Gaussian the recent promising advances in the reconciliation of Gaussian variables \cite{Jouguet:2011p5057} should find application in this protocol.  Alternatively Alice and Bob could also engage in a discretisation as outlined in \cite{Silberhorn:2002p149}.

\subsection{Eve Mutual Information}
We will be bounding Eve's information by calculating it for a Gaussian state with the measured CM.  The Gaussian entropies can be calculated using,
\eqn{\nn S(\rho) =  \sum_k (1+\lambda_k)\log(1+\lambda_k) - \lambda_k\log(\lambda_k)}
where $\lambda_k$ are the symplectic eigenvalues of $\gamma_\rho$ the covariance matrix of $\rho$.  These can be calculated in for example, \cite{Leverrier:2009p390} or in full generality \cite{Serafini:2006p5052}.
\subsubsection{Direct Reconciliation}
Eve's quantum information about Alice's data is given by,
\eqn{I(x_A:E) = S(\rho^{PS}_{E}) - S(\rho_E^{PS}|x_A)}
Given that Eve purifies the state $\rho_E^{PS} = \rho_{AB}^{PS}$.  The covariance matrix of $\rho_{AB}^{PS}$ can be calculated directly from Eq.(\ref{mom}) and is of the symmetric form Eq.(\ref{sym}) with entries
\eqn{a \ee V_A^{PS}+1\nn\\
b\ee  T^{PS}V_A^{PS} + T^{PS}\xi^{PS} + 1\nn\\
c\ee\nn \sqrt{T^{PS}((V_A^{PS})^2 + 2V_A^{PS}}) }
Alice then makes a heterodyne measurement introducing a vacuum mode C.  Although Bob homodynes and thus the second mode is unused it is not given to Eve. Eve still purifiest the state after measurement so $S(\rho_E^{PS}|x_A) = S(\rho_BC|x_A)$.  Utilising the symplectic beamsplitter transformation in Eq.(\ref{S}) we find the covariance matrix is,
\eq{\gamma_{BC|x_A} =\matfour{b-\frac{c^2}{a}& 0&\frac{\sqrt{2}c}{a+1}&0}{0&b&0&-\frac{c}{\sqrt{2}}}{\frac{\sqrt{2}c}{a+1}&0&\frac{2a}{a+1}&0}{0&-\frac{c}{\sqrt{2}}&0&\frac{a+1}{2}}}
\subsubsection{Reverse Reconciliation}
Eve's quantum information about Bob's data is given by,
\eqn{I(x_A:E) = S(\rho^{E}_{AB}) - S(\rho_E^{PS}|x_B)}
The first term is exactly the same as the direct reconciliation expression but the last is more complicated due to the additional operations in Bob's station.  Bob has 3 ancilla modes: a vacuum, mode $D$, for the two-mode squeezing interaction and an EPR of variance $N_B$, modes $G$ and $F$.  Before the measurement the full CM would be
\eqn{\gamma_{ABDFG} \nn = \matfive{a \hs\mathbb{I}_2&c\hs \sigma_z&Z&Z&Z}{c\hs\sigma_z&b\hs\mathbb{I}_2&Z&Z&Z}{Z&Z&\mathbb{I}_2&Z&Z}{Z&Z&Z&N_B\hs\mathbb{I}_2&\sqrt{N_B^2-1}\hs\sigma_z}{Z&Z&Z&\sqrt{N_B^2-1}\hs\sigma_z&N_B\hs\mathbb{I}_2} }
where is a $2\times2$ matrix of zeros.  The states interact via Eq.(\ref{gtrans}) and after Bob's measurement Eve still has the purification so $S(\rho_E^{PS}|x_B) = S(\rho_BDFG|x_B)$ where the covariance matrix is given by,
\eqn{\left(
\begin{array}{cccc}
 \gamma _A & \sigma _{\text{AD}} & \sigma _{\text{AF}} & \sigma _{\text{AG}} \\
 \sigma _{\text{AD}} & \gamma _D & \sigma _{\text{FD}} & Z \\
 \sigma _{\text{AF}} & \sigma _{\text{FD}} & \gamma _F & \sigma _{\text{FG}} \\
 \sigma _{\text{AG}} & Z & \sigma _{\text{FG}} & \gamma _G
\end{array}
\right)\nn}
where,
\begin{widetext}
\eqn{\nn \gamma_A\ee\nn \left(
\begin{array}{cc}
 a+\frac{c^2 \eta  T_B}{N_B \left(-1+T_B\right)-(-1+\eta +b \eta ) T_B} & 0 \\
 0 & a
\end{array}
\right)\\
\gamma_D \ee\nn\left(
\begin{array}{cc}
 b (-1+\eta )+\eta +\frac{(1+b)^2 (-1+\eta ) \eta  T_B}{N_B \left(-1+T_B\right)-(-1+\eta +b \eta ) T_B} & 0 \\
 0 & b (-1+\eta )+\eta 
\end{array}
\right)\\
\gamma_F \ee \nn \left(
\begin{array}{cc}
 \frac{(-1+\eta +b \eta ) N_B}{N_B+\left(-1+\eta +b \eta -N_B\right) T_B} & 0 \\
 0 & -1+\eta +b \eta +\left(1-(1+b) \eta +N_B\right) T_B
\end{array}
\right)\\
\gamma_G \ee\nn\left(
\begin{array}{cc}
 N_B-\frac{\left(-1+N_B^2\right) \left(-1+T_B\right)}{-N_B+\left(1-(1+b) \eta +N_B\right) T_B} & 0 \\
 0 & N_B
\end{array}
\right) \\
\sigma_{AD} \ee\nn\left(
\begin{array}{cc}
 -\frac{c \sqrt{-1+\eta } \left(N_B \left(-1+T_B\right)+T_B\right)}{N_B+\left(-1+\eta +b \eta -N_B\right) T_B} & 0 \\
 0 & c \sqrt{-1+\eta }
\end{array}
\right)\\
\sigma_{AF} \ee\nn\left(
\begin{array}{cc}
 c \sqrt{\eta } \left(-\sqrt{1-T_B}+\frac{\left(-1+\eta +b \eta -N_B\right) \sqrt{T_B} \sqrt{-\left(-1+T_B\right) T_B}}{N_B+\left(-1+\eta +b \eta -N_B\right) T_B}\right) & 0 \\
 0 & c \sqrt{\eta } \sqrt{1-T_B}
\end{array}
\right)\\
\sigma_{AG}\ee\nn\left(
\begin{array}{cc}
 \frac{c \sqrt{\eta } \sqrt{-1+N_B^2} \sqrt{-\left(-1+T_B\right) T_B}}{-N_B+\left(1-(1+b) \eta +N_B\right) T_B} & 0 \\
 0 & 0
\end{array}
\right)\\}
\end{widetext}

Finally the secret key rate must be multiplied by the fraction of the data kept in the post-selection process so the final key rate reads,
\eq{K = N(\beta I(a:b) - I(E:X)), \hs X \in \{a,b\}}
for direct and reverse reconciliation.
\end{document}